\documentclass[a4paper,12pt]{report}

\usepackage{amsmath}
\usepackage{comment}
[
        a4paper,% other options: a3paper, a5paper, etc
        left=3cm,
        right=3cm,
        top=1.5cm,
        bottom=1.5cm,
]
\usepackage{amsmath,amssymb}
\usepackage{amsmath}
\usepackage{textcomp}
\usepackage{wasysym}
\usepackage{braket}
\usepackage{graphicx}
\usepackage{xcolor}
\usepackage{caption}
\usepackage{subcaption}
\usepackage{float}
\usepackage{cite}
\usepackage{hyperref}
\usepackage{lipsum}
\usepackage{breqn}
\usepackage{siunitx}
\usepackage{pdfpages}

\begin{document}
\titlepage
\vspace{1.5cm}
\begin{center}
{\large \bf {Shape Transitions in Network Model of Active Elastic Shells}}\\
%\vspace{1.5cm}

\vspace{0.5cm}
Ajoy Maji\\
Faculty of Biomedical Engineering, Technion-Israel Institute of Technology, 32000 Haifa, Israel\\
\vspace{0.8cm}
Kinjal Dasbiswas\\
Department of Physics, University of California, Merced, Merced, CA 95343, USA\\
\vspace{0.8cm}
     
 Yitzhak Rabin\\
Department of Physics and Institute of Nanotechnology and Advanced Materials,
Bar-Ilan University, Ramat-Gan 5290002, Israel

\end{center}

\begin{abstract}
    Morphogenesis involves the transformation of initially simple shapes, such as multicellular spheroids, into more complex $3D$ shapes. These shape changes are governed by mechanical forces including molecular motor-generated forces as well as hydrostatic fluid pressure, both of which are actively regulated in living matter through mechano-chemical feedback. Inspired by autonomous, biophysical shape change, such as occurring in the model organism hydra, we introduce a minimal, active, elastic model featuring a network of springs in a globe-like spherical shell geometry.  In this model there is coupling between activity and the shape of  the shell: if the local curvature of a filament represented by a spring falls below a critical value, its elastic constant is actively changed. This results in deformation of the springs that changes the shape of the shell. By combining excitation of springs and pressure regulation, we show that the shell undergoes a transition from spheroidal to either elongated ellipsoidal or a different spheroidal shape, depending on pressure. There exists a critical pressure at which there is an abrupt change from ellipsoids to spheroids, showing that pressure is potentially a sensitive switch for material shape.  We thus offer biologically inspired design principles for autonomous shape transitions in active elastic shells.
\end{abstract}

\section{Introduction:}
Shape change driven by active mechanical forces is the hallmark of living matter during morphogenesis \cite{thompson1992}. How an embryo transforms itself from an initial spheroidal assembly of cells to a complex 3D shape in a programmed and robust manner is a central question in developmental biology \cite{forgacs2005}.  The robustness of biological pattern formation is thought to arise from the feedback loops between complex chemical signaling patterned by genetics, and mechanical forces \cite{howard2011, dasbiswas2016mechanobiological,hannezo2019mechanochemical}.  In animal cells, these forces are typically generated by myosin II molecular motors that transduce chemical energy (ATP) to mechanical work \cite{howard} in the cytoskeleton \cite{lecuit2007}:  a cross-linked viscoelastic network of fibers that constitutes the structural framework of the cell \cite{gardel2015}. Additionally, mechanical forces arising from osmotic pressure also determine tissue size and shape \cite{adar2020active,ref_5}. Whether active shape transitions seen in living matter can be realized in synthetic matter remains a key question for materials scientists.

% Active solids and bioinspired shape morphing materials,,  summary of recent models
Living soft matter, ranging from cytoskeletal assemblies to multicellular tissue, is typically considered as a viscoelastic gel which resists shear deformations \cite{fabry2011} under the action of internally driven molecular motor-generated mechanical forces \cite{prost2015active, bernheim2018living,hemingway2015active}. While at long time scales they can self-organize through active flows \cite{kruse2005generic,marchetti2013hydrodynamics}, at  short time scales relative to the cytoskeletal remodeling times, such materials behave as \emph{active solids} with a well-defined reference state about which elastic deformations can occur \cite{ideses2018,salbreux2017mechanics,turlier2014furrow,banerjee2011generic} . One may take inspiration from these examples of elastic active matter in biology \cite{burla2019mechanical} to design self-actuating metamaterials that comprise elastically coupled active units \cite{baconnier2022selective}.  These may exhibit collective excitations triggered by the nonlinear feedback between elasticity and internal active forces, leading to a design principle for self-organized and autonomously shape changing materials \cite{bertoldi2017flexible}.  Thin elastic shells, in particular, are highly deformable and susceptible to stimuli-responsive shape changes, including instability to buckling \cite{paulose2013}, and are therefore ideal candidates for creating shape-morphing smart elastic materials \cite{Holmes2019}. Recent bio-inspired models for programming shape change in elastic shells through mechanochemical control of active forces exhibit rich pattern formation phenomena such as wave propagation and oscillatory dynamics \cite{Miller2018, levin2020self,li2021chemically,yin2022three} as well as multiple classes of stationary wrinkle patterns \cite{ref_3, zakharov2021modeling}. 

Apart from chemical regulation that is natural in biology, geometric cues such as curvature \cite{stoop2015curvature,cao2008self} and physical cues such as the elastic properties of an underlying substrate \cite{binysh2022active} can direct shape change of an active elastic shell. Similarly in fluid surfaces, the coupling of active forces and fluid flows to surface geometry also leads to self-organized shape change and patterning \cite{mietke2019self, morris2019, fovsnarivc2019theoretical}.  In this work, motivated by morphogenetic shape changes in hydra, we ask how a pressurized elastic shell that is initially spheroidal (such as a monolayer of epithelial cells around a fluid-fluid lumen in the ``blastula'' stage of an embryo \cite{forgacs2005}) can break symmetry and elongate along a pre-selected axis to form an ellipsoidal shape under internally generated active forces. In synthetic materials, such shape changes may be programmed in through anisotropic gel swelling controlled by aligned fibers or liquid crystal elastomers \cite{gladman2016biomimetic, white2015}.  In biological systems, such as the model organism hydra, pre-established actomyosin  fibers provide a natural orientational cue to motor-generated active stresses. Defects in this orientational ordering have been shown to act as organizational centers of shape change  \cite{maroudas2021topological}. Motivated by the body plan of hydra, which features a bilayer of orthogonally arranged parallel arrays of muscle-like actin fibers \cite{ref_2}, we introduce a ``globe''-like geometry with elastic springs arranged along discrete lines of latitude and longitude. Such a  model geometry then features pre-existing bidirectional orientational order with defects at the poles, but does not presume the presence of nematic disclinations in the orientation of filaments comprising the elastic material of the shell \cite{pearce2020defect}.

Here, we show with two minimal model ingredients, 1. curvature-dependent active elastic forces, and 2. hydrostatic pressure regulation, that an initially spherical elastic shell can autonomously transition to elongated shapes. A central feature of all cases we present is the presence of a critical hydrostatic pressure at which the shell exhibits a discontinuous transition between spheroidal and elongated ellipsoidal shapes. The paper is organized as follows. In the \emph{Materials and Methods} section, we introduce the key ingredients of the elastic shell model: the ``globe''-like network geometry, the hydrostatic and elastic forces, the molecular dynamics method and the active excitation of springs implemented as curvature-dependent modification of spring stiffness. In the \emph{Results} section, we present the initial distribution of curvatures of elastic springs of the pressurized shell and discuss the three models of active excitation we implemented. Models 1 and 2 correspond to irreversible and reversible excitations of springs below a certain threshold curvature, respectively and give rise to qualitatively similar steady state behaviors. In both models active excitation of springs leads to a transition from the initial spheroidal shape to elongated ellipsoids at low pressures and to inflated spheroidal shapes at high pressures (with an abrupt transition between ellipsoids and spheroids at some critical value of the pressure).  In model 3, the springs are excited in a reversible fashion, both above (spring constants increase) and below (spring constants decrease) the curvature threshold. The effects of hydrostatic pressure are mquite complex in this model: while shrunken oblate spheroids form at low pressures, they change continuously with increasing pressure into slightly prolate ellipsoids. At some pressure a discontinuous shape and area transition takes place  appear at high pressures. In each case, we track the steady state size (surface area) and shape (aspect ratio) of the shells and provide intuitive understanding of these quantities through an analysis of the distribution of spring curvatures and the sequence of excitation events. We conclude by discussing the results in the context of  actively shape-changing biological and synthetic materials.

\section{Materials and Methods}

\subsubsection*{Geometry} 
\begin{figure}[H]
\centering
\includegraphics[width=.6\linewidth]{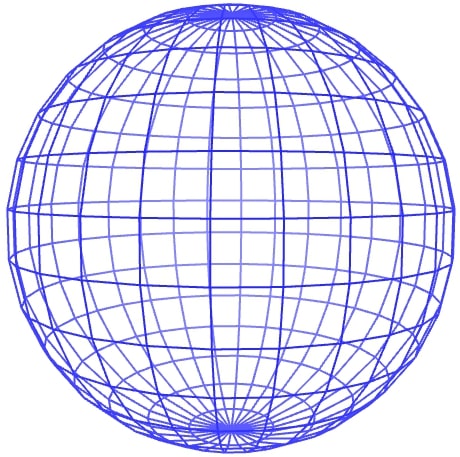}
\caption{The model network of the elastic shell in its initial spherically symmetric state before application of hydrostatic pressure and active excitations. Each segment along the lines of latitude or longitude is a nonlinear elastic spring which undergoes active modification of its spring constant according to its local curvature.}
\label{model}
\end{figure}

In the absence of externally imposed forces, the elastic system is represented by a connected network of non-linear springs that lie along the longitudinal and latitudinal circles on a sphere (globe) of radius $R=9$ as shown in Figure \ref{model}. 
The positions of the vertices of the network are given by their polar and azimuthal angles $\theta$ and $\phi$ in the spherical coordinate system, that vary in the intervals $[0,\pi]$ and $[0,2\pi]$, respectively. The entire range of $\theta$ is divided into $15$ equal intervals of size $\pi/15$ each and range of $\phi$ is divided into $28$ equal intervals of size $2\pi/28$. This construction yields $14$ latitudinal and $14$ longitudinal circles whose intersections yield $392$ vertices; two additional vertices that correspond to the poles of the globe are formed at the intersections of the longitudinal circles and can be considered as defects of the oriented structure. Note that since $28$ equally-spaced meridians (lines that connect the north and the south poles) meet at each pole, the network is invariant under rotation by any integer multiple of $2\pi/28$ about the symmetry axis (the line that connects the two poles). The $394$ vertices of the network are connected to each other by $812$ non-linear springs that form the elastic network whose structure ($364$ quadrilaterals and $28+28=56$ polar triangles) is maintained under arbitrary deformations.

\subsubsection*{Hydrostatic Forces} In order to introduce forces due to hydrostatic pressure that swells the network, we first complete the triangulation of the surface by dividing each of the $364$ quadrilaterals into $4$ virtual triangles  by locating the center of mass of each quadrilateral and joining each of its vertices to the center of mass (note that the virtual lines joining the  $4$ vertices of the quadrilateral to its center of mass do not correspond to elastic springs). This construction yields $1456$ non-polar triangles, in addition to the $56$ polar ones (see Figure S1 in Supplementary Information).
 Following the methods of reference \cite{ref_1} we define the hydrostatic force $\Vec{F}^h_{l}$ on vertex $l$ as the product of the pressure $p$ and the vector sum of the areas of the $n$ triangles that meet at this vertex. Here $n=28$ for each of the two poles,  $n=6$ for vertices that are connected by springs to the poles, and $n=8$ for all other vertices:
 
\begin{equation}
    \Vec{F}^h_{l}=p\sum_{i=1}^{n} \Vec{a_{i}}
\end{equation}
where $\Vec{a_{i}}$ is the area vector of the $i^{th}$ triangle.
The total area of the network is the sum of the areas of all triangles,
\begin{equation}
    A=\sum_{i=1}^{1456+56}a_{i}
\end{equation}

\subsubsection*{Elastic Forces} The elastic force on vertex $l$ due to the spring connecting it to vertex $q$ is given by 
\begin{equation}
    \Vec{F}^s_{lq}=-K_{lq} \frac{\Vec{r_l}-\Vec{r_q}}{|\Vec{r_l}-\Vec{r_q}|} (|\Vec{r_l}-\Vec{r_q}|-l^{eq}_{lq})^3
\end{equation}
Here $K_{lq}$ and $l^{eq}_{lq}$ are the spring constant and equilibrium length of the $lq$ spring, respectively. We use non-linear elastic springs since linear springs cannot stabilize the system against runaway expansion in the presence of hydrostatic pressure \cite{ref_1}. In the following we set $K_{lq}=K_{eq}=1$ for all the springs in the system (in the absence of activity). The equilibrium lengths of all springs are defined as their lengths in the ground state of the network (the sphere shown in Figure \ref{model}) in the absence of hydrostatic pressure, $p=0$. In other words, these equilibrium lengths are given by the lengths of the corresponding sides of the $56$ polar triangles and $364$ quadrilaterals shown in Figure \ref{model}. The total elastic force on vertex $l$ is the vector sum of the contributions from all the springs that connect this vertex to its immediate neighbors:
 \begin{equation}
    \vec{F}^s_{l}=\sum_{q=1}^{n} \Vec{F}^s_{lq}
\end{equation}

\subsubsection*{Equations of Motion} In addition to elastic and hydrostatic forces, there is a frictional force $-\zeta \vec{v}_{l}(t)$ acting on each moving vertex which is proportional to its instantaneous velocity $\vec{v}_{l}(t)$.  Here  $\zeta$ is the friction coefficient.

 Hence the equation of motion for the $l^{th}$ vertex is,
    \begin{equation}
        m\dot{\Vec{v}}_{l}(t)=-\zeta \Vec{v}_{l}(t)+\Vec{F}_{l}(t)
        \label{equ_motion}
    \end{equation}
where $\Vec{F}_{l}=\Vec{F}^h_{l}+\Vec{F}^s_{l}$. We set the mass $m$ of each vertex to unity and its friction coefficient $\zeta$ to $6$ and therefore the damping time is $\tau_{damp}=m/\zeta=1/6$ in time units used throughout our simulation (unless otherwise specified, the Molecular Dynamics (MD) time step is $\Delta t=10^{-3}$ in these time units). The equations of motion are solved using an improved algorithm (see SI) for simultaneously updating the velocity and the position of each vertex. Note that the velocities of all vertices  vanish in steady state.

Since the average coordination number of our network $z=(392\times 4+2\times 28)/394=4.12$ is less than $2d$, where $d=3$ is the dimension of space in which the network is embedded, the Maxwell stability criterion implies that the ground state of our network is unstable against shear \cite{Maxwell1864, ref_9}. In order to stabilize it we introduce hydrostatic pressure $p_0$ whose role is similar to that of osmotic pressure that stabilizes three dimensional gels against collapse (there it originates from excluded volume forces between the polymers in the network). In the following we take $p_0=0.1$ which results in about a four-fold increase in the surface area of the initially spherical shell (from $1008$ to $3943$). The steady state configuration of the network in the presence of this hydrostatic pressure (in the absence of active forces) is that of an oblate spheroid resembling the Earth (a snapshot of this configuration is shown in Figure S2 in SI).  We note here that fibrous biopolymer networks, while similarly under-coordinated, are mechanically stabilized by the bending stiffness of the constitutent fibers \cite{broedersz2014}. Since we are interested in exploring shape changes in a minimal elastic network with central force springs alone, rather than in a specific model for an actin fiber network, we ignore here the additional effects of bending stiffness.

\subsubsection*{Active Forces}
Following our previous work \cite{ref_1} we introduce active forces by changing the elastic parameters of the network, i.e., change the spring constants from their equilibrium value $K_{eq}$, to their value in the active state $K$ (an alternative approach which yields qualitatively similar results is to change the equilibrium lengths of the springs \cite{ref_1}). Since decreasing (increasing) $K$ results in elongation (contraction) of the springs, such active excitations simulate the action of a muscle (reminiscent of the hydra tissue which is a muscle made of cells that actively
stretch and compress). On a microscopic level such changes involve some underlying chemical mechanism, e.g., a mechanochemical coupling between the local concentration of a morphogen-like diffusing chemical signal (such as calcium ions \cite{ref_2}) and the contractility of the springs that represent the elastic system (e.g., actomyosin fibers) \cite{ref_3}. We will adopt here a simplified version of the model of reference \cite{ref_3} in which the concentration of the chemical signal induces a sharp transition between a low and a high contractility state. In the present case, the contractility depends on the local curvature $\chi$ of filaments that run along the longitudinal and the latitudinal circles of the elastic shell. In this two-state model the springs become activated (their spring constants are changed) once this local curvature falls below or exceeds some critical value $\chi_c$. The underlying assumption is that the chemical signaling kinetics is fast and adapts immediately to the local curvature, which acts as the global patterning field for active forces. 

Since the springs are oriented along longitudinal/latitudinal circles that lie along the principal axes of curvature of the surface, their curvatures can be identified with the corresponding principal curvatures of the surface at their location. We define the curvature of a spring as follows. Any three connected longitudinal/latitudinal springs can be considered as a segments of a polymer that lie along the corresponding longitude/latitude.  Lets denote by $\theta_i$ and $\theta_{i+1}$ the angles between the tangent to the $i^{th}$ spring and the tangents to its neighbors along the longitude/latitude  (see  Figure S2). The dimensionless curvatures $\Tilde{\chi}=\chi\cdot l_i$ at the two neighboring vertices connected by spring $i$ ($l_i$ is the length of the  spring) are given by \cite{ref_4} $\Tilde{\chi}(v_i)=\sqrt{2(1-cos\theta_i)}\approx \theta_i$  and $\Tilde{\chi}(v_{i+1})=\sqrt{2(1-cos\theta_{i+1})}\approx \theta_{i+1}$. Averaging the two expressions yields the curvature of spring $i$, 
\begin{equation*}
\chi_i=\frac{\theta_i+\theta_{i+1}}{2l_i}
\end{equation*}

So far, we only considered curvature-dependent active excitations of the elastic springs of the network. For reasons that will become clear later in this work, we will allow for the possibility of changing the hydrostatic pressure inside the elastic shell, e.g., by active control of the transport of ions through the thin shell \cite{ref_5}.

\section*{Results}
\label{Results}
Consider a network in steady state in presence of hydrostatic pressure, $p_0=0.1$. This steady state is prepared by starting from the ground state configuration at $p=0$, raising the pressure to $p_0=0.1$ and allowing the network to expand to a steady state configuration.

\subsection{Curvature Distributions}
\begin{figure}[h!]
\centering
\includegraphics[width=.9\linewidth]{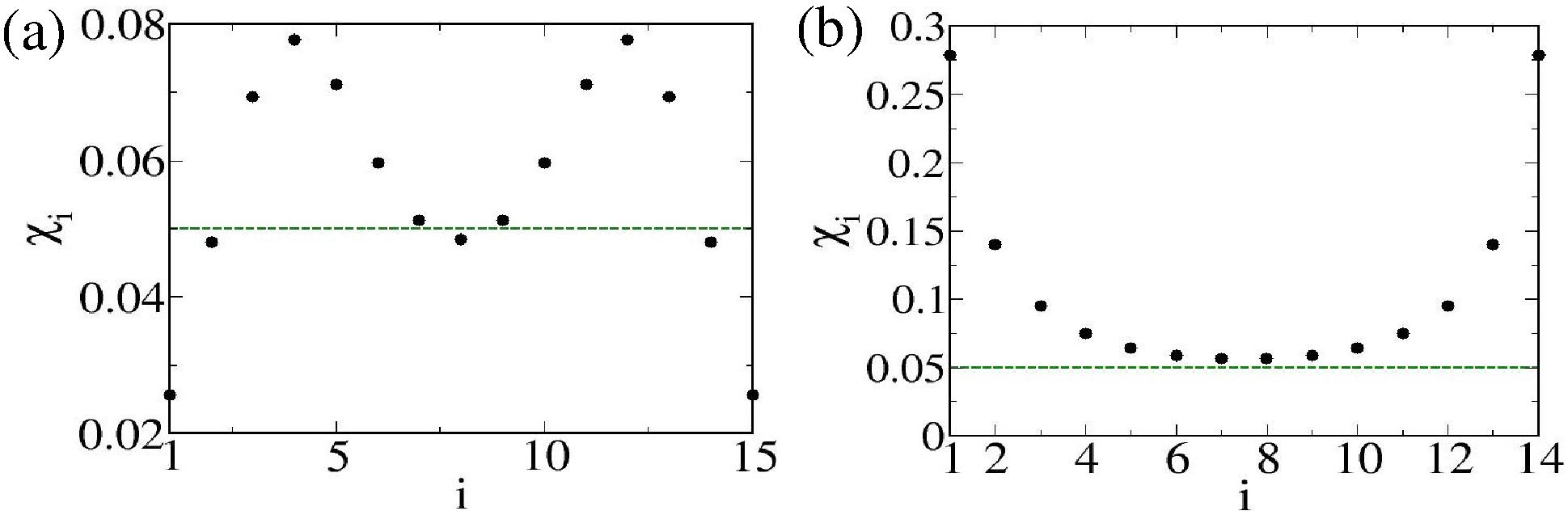}
\caption{Steady state curvatures at $p_0=0.1$ of (a) $15$ springs that lie along a single longitudinal circle, from south pole to the north pole, and (b) $14$ springs each of which lies on a different latitudinal circle, from one adjacent to the south pole to that adjacent to the north pole. The critical curvature $\chi_c=0.05$ is represented by the green horizontal line}
\label{fig_2}
\end{figure}

We proceed to calculate the curvatures $\chi_i$ of the springs in the steady state with $p_0=0.1$, in the absence of active excitations (all spring constants have their equilibrium values, $K_{eq}=1$). In Figure \ref{fig_2} we present a histogram of curvatures of the springs in the network. Since all longitudinal circles are equivalent by symmetry (see Figure \ref{model}), we consider springs that lie on the same longitudinal circle and label them by increasing values of $i$ starting from the south pole ($i=1$) and going to the north pole ($i=15$). Since all springs along a single latitudinal circle are equivalent by symmetry, we choose a single spring from each latitudinal circle and label them by index $i$ that goes from $i=1$ for the latitudinal circle closest to the south pole, to $i=14$ for the latitudinal circle adjacent to the north pole.
Note that while all longitudinal springs have the same curvature in the ground state configuration (in the absence of pressure the network lies on a sphere and all great circles have equal lengths), this degeneracy is lifted by the application of hydrostatic pressure that deforms the sphere into an oblate spheroidal shape 
(see Figure \ref{fig_2}a and Figure S3 in SI). The curvature histogram of longitudinal springs has two symmetric minima at the poles and a local minimum at the equator (there is north-south symmetry with respect to the equatorial plane).  The histogram of curvatures of the latitudinal springs  is nearly parabolic, with curvatures increasing as one moves from the equator towards the poles ( see Figure \ref{fig_2}b), reflecting the decrease of the diameters of the corresponding latitudinal circles. 

\subsection{Models of Activation}
In the following we will explore several mechanisms of curvature-dependent excitation of the elastic network. Our interest in the problem arose from recent studies of development of reconstituted hydra in which the organism undergoes a morphology transition from spheroidal to elongated shape. In particular, the construction of a network of springs that lie along the longitudinal and latitudinal circles was modeled after the arrangement of actomyosin fibers that run through the two epithelial layers of the hydra spheroid. While this arrangement breaks rotational symmetry and defines an axis of elongation, an active excitation that would simply reduce the spring constants of all springs would lead to isotropic expansion and would not generate elongation. 
%Previously we found that excitations of randomly chosen springs will introduce random deformations of the spheroid but will not result in its elongation \cite{ref_1}. 
This raises the question: how can one obtain elongation in our simple model of active elastic shells? Clearly,  anisotropic deformation would result if one could elongate at least some of the longitudinal springs (by decreasing their spring constants) while not deforming the latitudinal ones. This suggests that if we were to introduce a critical curvature $\chi_c$ that lies below the bottom of the spectrum of curvatures of latitudinal springs in Figure \ref{fig_2}, and excite only springs with $\chi<\chi_c$, only longitudinal springs will be elongated and the spheroid will deform into an elongated ellipsoid. The surprising consequences of this simple idea are explored in the following.

\subsubsection*{Models $1$ and $2$. Irreversible and reversible activation: spheroid to ellipsoid transition}
Here and in the following we take $\chi_c=0.05$, a value which lies slightly below the lower cutoff of the curvature spectrum of latitudinal springs (see Figure \ref{fig_2}). Starting from the steady state spheroidal configuration of the network (Figure S2 in SI), at time $t=500$ we introduce active excitations i.e., change the spring constants of all springs with $\chi_{i}<0.05$ from $K_{eq}=1$ to $K=0.05$, and simultaneously change the pressure from its initial value of $p_0=0.1$ to some final value $p$. In model $1$ the change of spring constants is assumed to be irreversible, i.e., the excitation of a spring does not relax even if its curvature increases again above $\chi_c$, in the process of deformation of the network. 
This constraint is removed in model $2$ in which activation is reversible i.e., if the curvature of a spring increases above $\chi_c$, the corresponding spring constant instantaneously recovers its equilibrium value, $K_{eq}=1$. The sequence of activation and relaxation events can be repeated again and again, depending on the instantaneous local curvature, until a steady state is reached in which the shape of the network and the spring constants no longer change. 

Contrary to the expectation that the introduction of relaxation of activated springs would have a major effect on the dynamics, we find that models $1$ and $2$ yield qualitatively similar results. Therefore, in the following we present the results for the irreversible case only (model 1); the corresponding results for the reversible case are relegated to the SI (see section Model $2$, Figures $S4-S6$). 
%The main difference is that the transition takes place at lower pressure and the aspect ratio of the resulting prolate ellipsoid is higher in model $1$ compared to model $2$.

Returning to model $1$, we expect that if the pressure is left unchanged or is increased compared to its value before the onset of excitation, the $20$-fold reduction of the spring constants of the excited springs will result in dramatic swelling of the network. 

\begin{figure}[h!]
\centering
\includegraphics[width=1.0\linewidth]{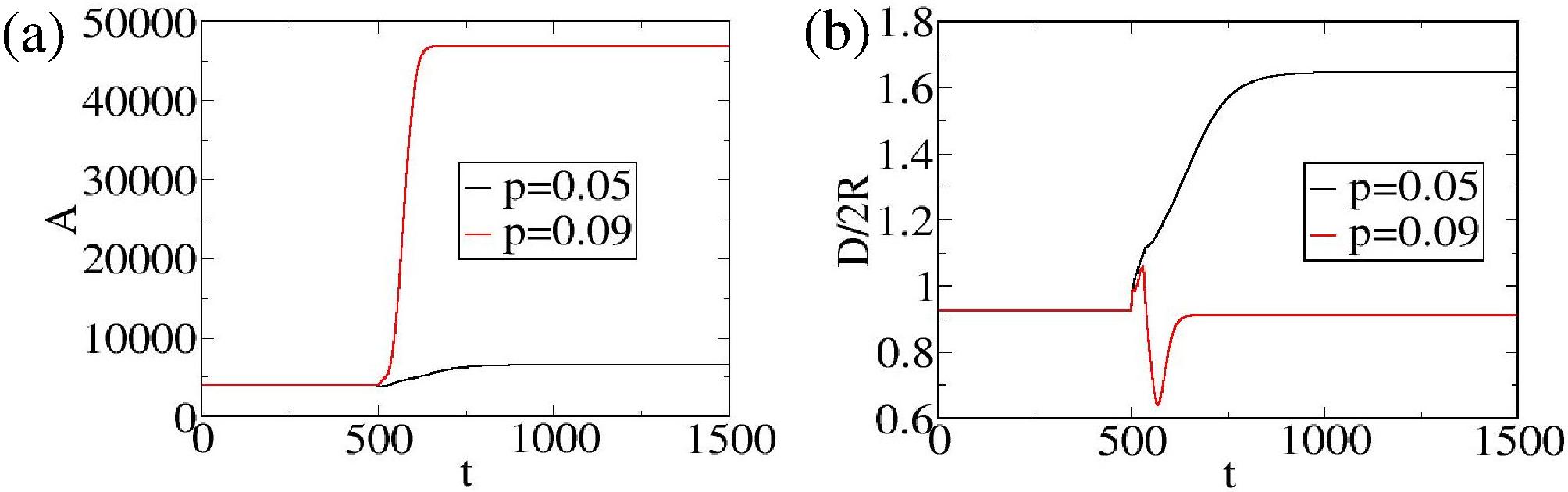}
\caption{Model 1: (a) Area A and (b) aspect ratio ($D/2R$) are plotted as a function of time for two different values of pressure, $p=0.05$ (in black) and $p=0.09$ (in red). In both cases, the springs below a curvature threshold are activated as well as pressure is reduced from its initial value $p_{0} = 0.1$, at time $t=500$.}
\label{Model_1_1}
\end{figure}

\noindent This effect is clearly observed in Figure \ref{Model_1_1}a for $p=0.09$ (which is close to the initial value $p_0=0.1$): the total area $A$ (red curve) increases with time and saturates at a value that is nearly an order of magnitude higher than the initial area. However, if the pressure is lowered well below its initial value, the weakening of the spring constants is balanced by the decrease of the pressure and only a minor increase of the area is observed in steady state (see black curve that corresponds to $p=0.05$ in Figure \ref{Model_1_1}a). The reduction of hydrostatic pressure inside the elastic shell can be achieved by active control of osmotic pressure. Indeed, a mechanism based on osmoregulation, i.e, on the ability of a living organism to control the osmotic pressure inside it through active ion transport \cite{adar2020active}, was invoked to explain the observed shape oscillations in a regenerating hydra \cite{ref_5}. 

In Figure \ref{Model_1_1}b we plot the time series of the aspect ratio $D/2R$, where $D$ is the distance between the poles and $R$ is the radius at the equator (see Figure $S3$ for an illustration).  Note that $D/2R=1$ corresponds to a sphere, $D/2R>1$ to a prolate and $D/2R<1$ to oblate ellipsoid. For $p=0.09$ the aspect ratio first increases with time and then drops sharply, increases again and eventually saturates at a value below unity corresponding to a spheroidal/oblate ellipsoidal configuration (red curve in Figure \ref{Model_1_1}b and Movie M1 in SI). Upon some reflection we conclude that initially only longitudinal springs are excited and a prolate ellipsoidal shape is transiently formed (the aspect ratio increases). However, as swelling proceeds and the surface area of the elastic shell further increases, the curvature of many latitudinal  springs falls below $\chi_c$ and they become activated as well (their spring constants drop from $1$ to $0.05$). As a result, the network expands in the transverse direction (the aspect ratio decreases) and assumes a spheroidal/oblate ellipsoidal shape in the final steady state (the intermediate dip in the aspect ratio is due to inertial effects). For smaller pressure, $p=0.05$, the change in the area is small and since the curvatures of latitudinal springs remain above the critical value, their spring constants remain unchanged ($K=1$) during the process of deformation. Since only longitudinal springs are elongated,  the aspect ratio increases monotonically with time and saturates at a value of about 1.65 that corresponds to a prolate ellipsoidal shape (black curve in Figure \ref{Model_1_1}b and movie $M2$ in SI).

\begin{figure}[H]
\centering
\includegraphics[width=0.8\linewidth]{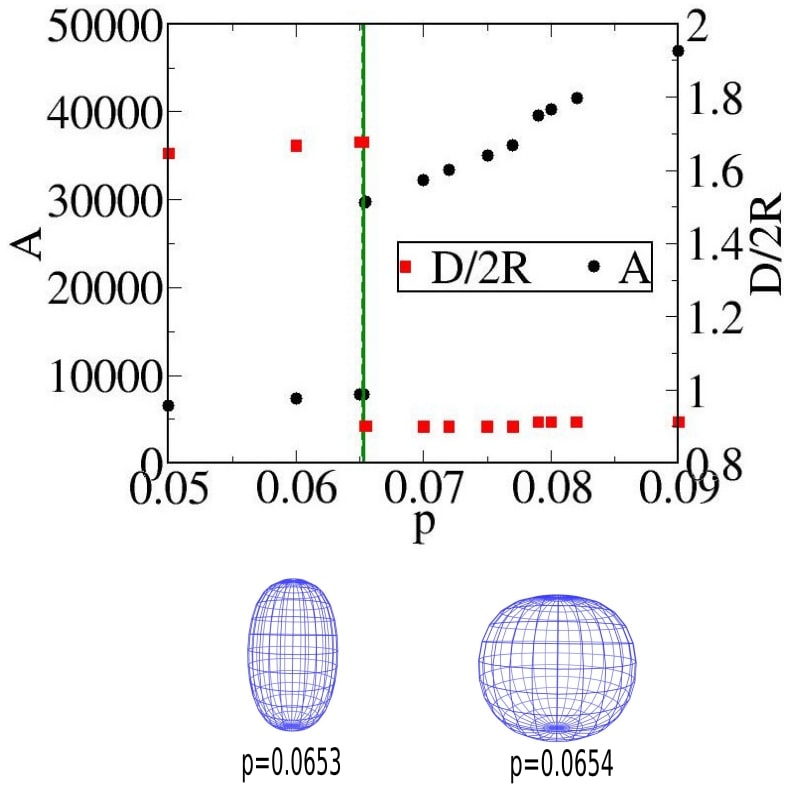}
\caption{Model 1: Steady state area A (black dots) and aspect ratio D/2R (red squares) are plotted as a function of pressure $p$. The dashed green vertical line shows the location of the discontinuous transition. Snapshots of the steady state shapes at  pressures $p=0.0653$ and $p=0.0654$, on both sides of the transition, are shown in the lower panel (the snapshots are not to scale).}
\label{Model_1_2}
\end{figure}

In Figure \ref{Model_1_2} we present the values of the area (black dots) and the aspect ratio (red squares) in the final steady state of the actively excited network, as a function of pressure in the range $0.05\leq p\leq 0.09$. A discontinuous transition from prolate ellipsoidal to spheroidal shape accompanied by a jump of the total area of the network, is observed at $p=p_c$ (shown by the dashed green vertical line in Figure \ref{Model_1_2}). Snapshots of the steady state shapes of the system on both sides of the transition, at $p=0.0653$ and $p=0.0654$, are shown in the lower panel. 
%(the latter case was computed at higher accuracy by decreasing the MD time step from $0.01$ to $0.001$) 

\begin{figure}[H]
\centering
\includegraphics[width=1.0\linewidth]{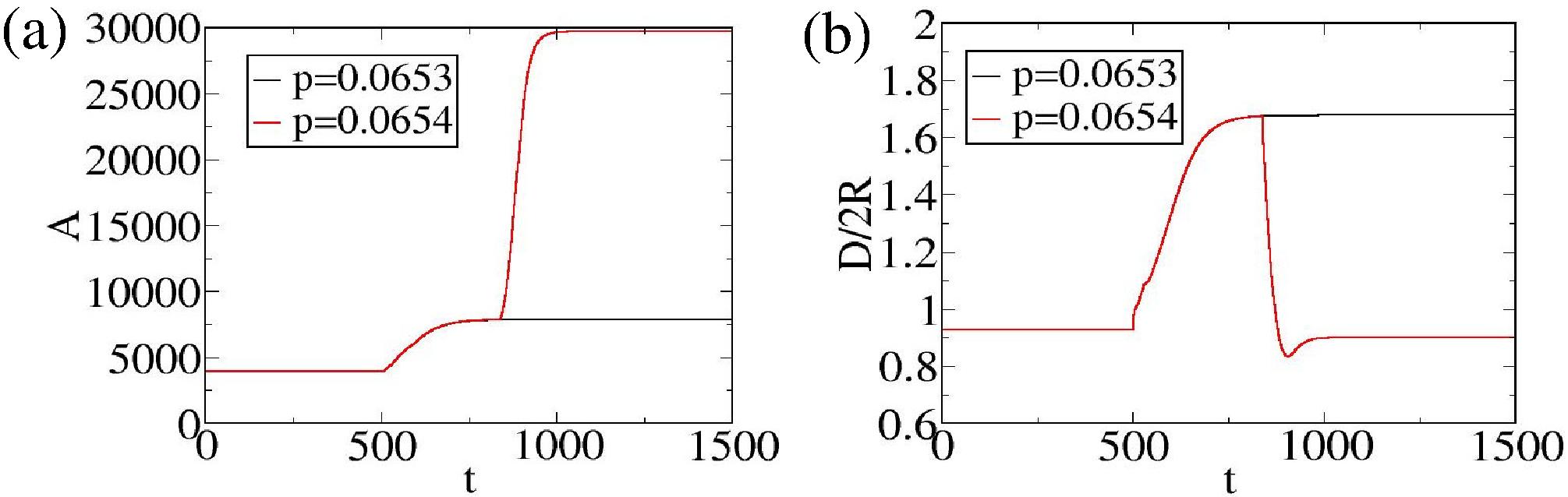}
\caption{(a) Area A and (b) aspect ratio D/2R are plotted as a function of time just below  (p=0.0653, in black) and just above (p=0.0654, in red) the transition.}
\label{Model_1_5}
\end{figure}
Inspection of the time histories of the area and the aspect ratio in Figures \ref{Model_1_5}a and b reveals an interesting dynamical asymmetry between the two sides of the transition. Below the transition, at $p=0.0653$, both the area and the aspect ratio increase monotonically with time until a prolate ellipsoid configuration is reached (movie M3 in SI). Above the transition, at $p=0.0654$, the initial increase of the volume and the aspect ratio is identical to that below it and a prolate ellipsoid is transiently formed. However, as the expansion continues, the system undergoes rapid swelling, the aspect ratio drops below unity, and a dramatically swollen oblate spheroidal shape results (movie M4 in SI). Similar dynamical asymmetry is observed away from the transition, in Figures \ref{Model_1_1}a and b  (compare the curves corresponding to $p=0.05$ and $p=0.09$ to those in Figures \ref{Model_1_5}a and b), but the time at which the dynamical histories diverge from each other is much shorter. 

\begin{figure}[h!]
\centering
\includegraphics[width=1.0\linewidth]{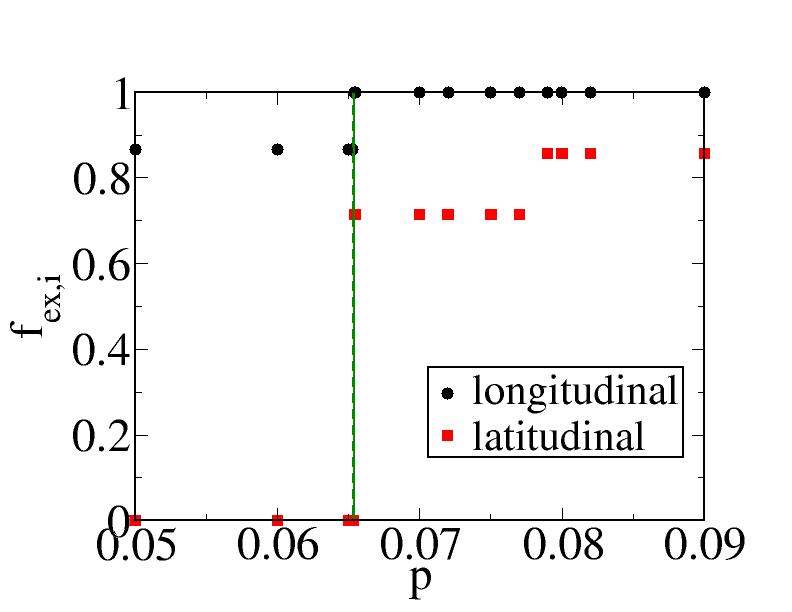}
\caption{Model 1: Fractions of excited springs in the final steady state, $f_{ex}$, for different pressures p}
\label{Model_1_3}
\end{figure}

 In order to understand the origin of this transition, in Figure \ref{Model_1_3} we plot the fractions of the excited latitudinal and longitudinal springs in the final steady state of the network, where the fraction $f_{ex,i}$ of excited springs of type $i$ is defined as the ratio of the number of excited springs of type $i$ to the total number of springs of that type. 

Inspection of Figures \ref{Model_1_2} and \ref{Model_1_3} shows that the prolate-oblate ellipsoid transition at pressure $p_c$ corresponds to a discontinuous jump between a low pressure  phase in which only longitudinal springs are excited, to a high pressure phase in which more than $65\%$ of the latitudinal springs are excited (a smaller jump is also observed at $p_c$ in the case of longitudinal  springs for which $f_{ex}$ increases from about $0.87$ to unity- see Figure \ref{Model_1_3}). For pressures below $p_c$, increasing the pressure increases both the fraction of activated longitudinal springs (the latitudinal springs are not activated - see Figure  \ref{Model_1_3}) and the surface area of the elastic shell (Figure  \ref{Model_1_2}). Since the swelling of the network reduces the curvatures of all springs, one reaches a pressure $p=p_c$ above which the curvatures of the latitudinal springs in the equatorial region (these have the smallest curvatures among latitudinal springs) become lower than $\chi_c$ and they are activated. As the reduction of their spring constants produces further expansion of the network, the curvatures of the springs are reduced and additional latitudinal springs become excited. This positive feedback between excitation of springs and expansion of the network gives rise to a runaway process that leads to the discontinuous transition observed in Figures \ref{Model_1_2} and \ref{Model_1_3} at $p=p_c$. Above the transition, all the longitudinal and most of the latitudinal springs are excited and an oblate spheroidal shape results (see lower panel in Figure  \ref{Model_1_2}).  

\begin{figure}[h!]
\centering
\includegraphics[width=1.0\linewidth]{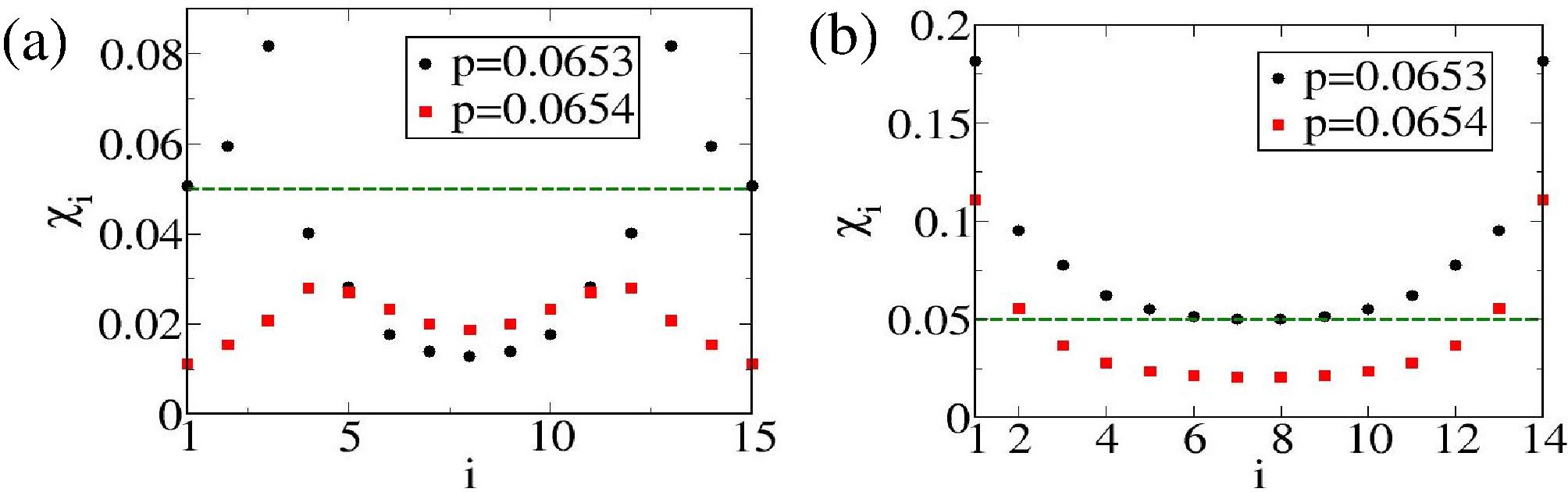}
\caption{Model 1: The steady state spectra of curvatures $\chi_i$ ($i$ is the spring index) of (a) latitudinal and (b) longitudinal springs, both below (at $p=0.0653$, in black) and above (at $p=0.0654$, in red) the transition. The horizontal broken green line shows the critical curvature, $\chi_c=0.05$.}
\label{Model_1_4}
\end{figure}

The discontinuous change of the spectrum of curvatures of latitudinal springs as the pressure is increased from below ($p=0.0653$) to above ($p=0.0654$) $p_c$, is demonstrated in Figures \ref{Model_1_4}a and \ref{Model_1_4}b. The critical curvature $\chi_c=0.05$ is shown by the horizontal broken green line in both figures.

\subsubsection*{Model $3$: All springs are excited:  spheroid to inflated elongated ellipsoid (spheroid to shrunken spheroid) transition above (below) critical pressure} 
So far, we only considered active excitations of subcritical springs, i.e., springs whose curvature was lower than $\chi_c$. We now proceed to the case in which all springs are excited, according to the following rule:   the elastic constants of springs whose instantaneous curvature is smaller than $\chi_c$ are reduced to $K^{<} < K_{eq}$, and those of springs whose curvature is larger than $\chi_c$ are increased to $K^{>} > K_{eq}$. Note that all springs are either stretched or compressed following the onset of excitation at $t=500$, when the springs are activated and the pressure is changed from its initial steady state value of $0.1$ to $p$. In the following, we take $K^{<}=0.2$ and $K^{>}=5$.  
%Because of numerical accuracy issues the MD time step was decreased from $10^{-2}$ used in simulations of models $1$ and $2$, to $10^{-4}$ (below which the results were independent of the time step used). 
Since the shape of the shell (and therefore the local principal curvatures) changes during deformation, springs can repeatedly change their elastic constants  until the network reaches the final steady state configuration. As the curvatures of most of the springs in the initial steady state lie above the critical value of $0.05$ (see Figure \ref{fig_2}), if the pressure is not changed from its initial value of $0.1$, the number of compressed springs exceeds that of stretched ones, leading to the shrinking of the network. The volume decrease will lead to further increase of local curvature and the resulting positive feedback will shrink the network even more until a steady state configuration is reached. 

\begin{figure}[h!]
\centering
\includegraphics[width=0.9\linewidth]{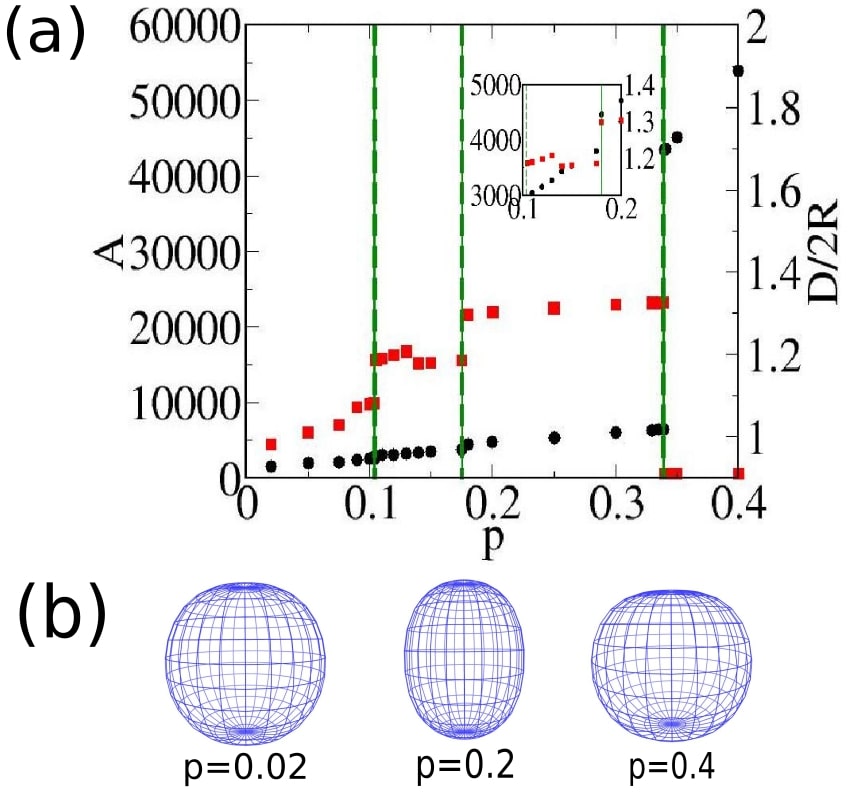}
\caption{Model $3$: (a) The final steady state values (in the presence of active excitations) of the total area (black dots) and the aspect ratio (red squares) are plotted as a function of pressure $p$.  The  vertical green lines in (a) show the pressures $p_{c1}$, $p_{c2}$ and $p_{c3}$ at which discontinuous transitions take place. An expanded view of the region $0.1\leq p\leq 0.2$ is shown in the inset. The spring constants in the active state are  are $K^{>}=5$ and  $K^{<}=0.2$. (b) Snapshots of  steady state configurations at three different pressures (the snapshots are not to scale).}
\label{model_c0.1}
\end{figure}

%If, at the onset of active excitation, the pressure $p$ is increased above its initial value of $0.1$, the final steady state area will increase as well and will eventually exceed the initial value of $\approx 4000$. 

In Figure \ref{model_c0.1}a we plot the area and the aspect ratio in steady state with active excitations, as a function of pressure in the range $0.02\le p\le 0.4$, for $K^{>}=5; K^{<}=0.2$. 
At the lowest pressures (below $p=0.05$) the steady state configuration corresponds to a shrunken oblate spheroid, with aspect ratio below unity. As pressure is increased towards the first transition at $0.104<p_{c1}<0.105$, both the area and the aspect ratio increase continuously and the oblate spheroid becomes a slightly prolate ellipsoid. Then, at $p_{c1}$ there is a discontinuous jump in the area and the aspect ratio as the ellipsoid becomes larger and more elongated. Inspection of Figure $S7a$ in the SI shows that this transition is associated with the change of curvature of some of the longitudinal bonds (located around the equator), from above to below $\chi_c$, that results in their spring constants dropping from $K^{>}=5$ to $K^{<}=0.2$ (the curvatures of all the latitudinal springs remain larger than $\chi_c$  (Figure $S7b$) and their spring constants remain at $K^{>}=5$). 

As pressure is increased above $p_{c1}$, the ellipsoid continues to grow in size but its aspect ratio exhibits complex non-monotonic behavior related to the non-trivial effect of pressure on the curvature distributions (see inset in Figure \ref{model_c0.1}a). In order to understand this behavior we note that as pressure increases from $0.13$ to $0.14$, the curvatures of longitudinal springs that are nearest neighbors to equatorial springs, decrease below $\chi_c$ (see Figure $S8a$ in SI) and their spring constants decrease from $5$ to $0.2$. At the same time the curvatures of springs that are next nearest neighbors to the equatorial ones increase above $\chi_c$ (see Figure $S8a$ in SI) and their spring constants change from $0.2$ to $5$ (the curvatures of all latitudinal springs remain above $\chi_c$ at both pressure values - see Figure $S8b$ in SI). The net effect of these curvature and spring constant variations as pressure is increased from $0.13$ to $0.14$, is to slightly decrease the aspect ratio (see inset in Figure \ref{model_c0.1}). This non-trivial behavior illustrates the complex nature of the response of the connected non-linear network to activity-induced changes of spring constants in our model.

Upon further increase in pressure, another discontinuous increase of the area and the aspect ratio is observed at $0.175<p_{c2}<0.18$ (see inset in Figure \ref{model_c0.1}a). Inspection of Figure $S9a$ in SI shows that this transition is associated with the change of curvature of springs situated between the poles and the equator (some of them decrease below $\chi_c$ and others increase above $\chi_c$). Another discontinuous change of area that increases by nearly an order of magnitude and of aspect ratio that drops below unity, is observed at 
$0.339<p_{c3}<0.34$. Inspection of Figure $S10b$ in  SI shows that this transition is associated with the curvature of most of the latitudinal springs  (except those on the polar circles) changing from above to below $\chi_c$ (above this transition all the longitudinal springs have curvatures below $\chi_c$ - see Figure $S10a$ in SI). The resulting reduction of the spring constants of these latitudinal springs gives rise to runaway radial expansion normal to the polar axis, via the same mechanism that has led to the elongated ellipsoid to oblate spheroid transition in models $1$ and $2$.

The re-entrant character of the transitions in model $3$ is clearly observed in the snapshots shown in Figure \ref{model_c0.1}b (the sizes are not to scale):  shrunken spheroids at lowest pressures, prolate ellipsoids at intermediate pressures inflated spheroids at highest pressures.

%\begin{figure}[h]
%\centering
%\includegraphics[width=1.0\linewidth]{fig_9.jpg}
%\caption{Model $3$:  Comparison of (a) longitudinal and (b) latitudinal spectra below ($p=0.104$) and above ($p=0.105$) the first discontinuous transition $p=p_{c1}$.}
%\label{curvature spectra near 1st transition}
%\end{figure}

\begin{figure}[h!]
\centering
\includegraphics[width=1.0\linewidth]{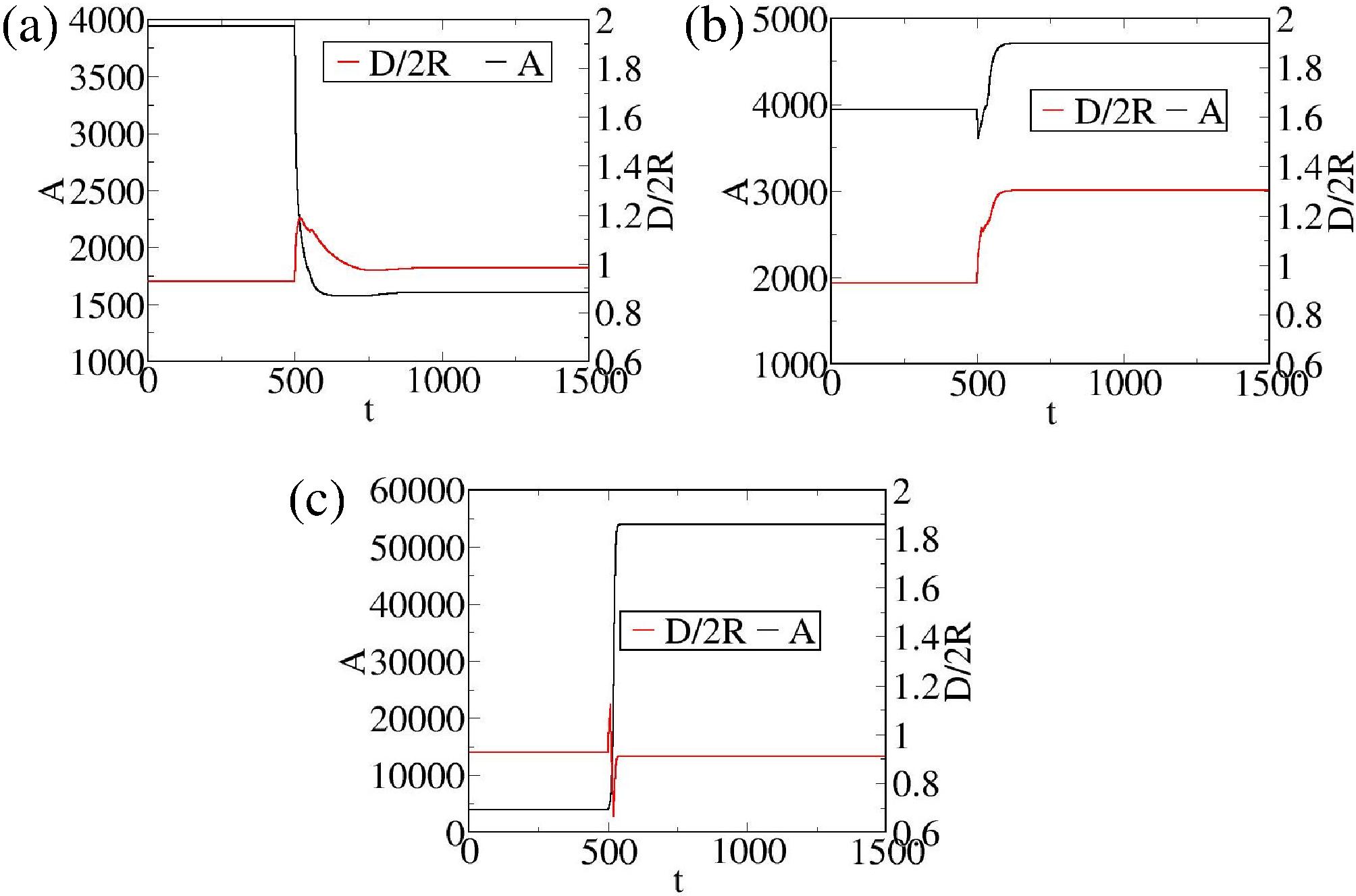}
\caption{Model $3$: Aspect ratio and area are plotted versus time for (a) $p=0.02$, (b) $p=0.2$ and (c) $p=0.4$.}
\label{time_history}
\end{figure}

The complex kinetics of the shape (area and aspect ratio) for $p=0.02$, $0.2$ and $0.4$ are shown in Figures \ref{time_history}(a), (b) and (c) (see also movies $M5$, $M6$ and $M7$ in SI). 
\noindent As pressure is reduced from $p_0=0.1$ to $p=0.02$, the area of the spheroid decreases and its curvature increases. Since all the latitudinal springs and some of the longitudinal ones have $\chi>\chi_c$, activation leads to net compression of the network as the corresponding spring constants increase to $K^>=5$ and the shrinking of the area becomes a runaway process (see Figure \ref{time_history}a) during which the area of the spheroid drops to about half of its initial value. This contraction is accompanied by transient increase of the aspect ratio, due to the compression of latitudinal springs (reminiscent of biaxial compression of materials with positive Poisson ratio), but the effect is rather weak and the elastic shell relaxes to an oblate spheroidal shape. At $p=0.2$ the initial contraction of longitudinal springs with $\chi>\chi_c$ is overcome by the increased pressure and monotonic increase of the total area with time up to saturation is observed (Figure \ref{time_history}b). The aspect ratio first dips (due to contraction of some longitudinal springs) and then increases monotonically with time up to saturation, as expansion reduces the local curvatures from above critical to sub-critical values, and the spring constants of the affected springs change from $K^{>}=5$ to $K^{<}=0.2$ (note that because of the shape of the curvature spectrum, Figure \ref{fig_2}, expansion affects mostly longitudinal springs). At $p=0.4$ the initial area increase affects mostly longitudinal springs (the curvature of additional longitudinal springs drops below $\chi_c$), their spring constants decrease and the aspect ratio increases. As expansion continues, the curvature of additional latitudinal springs decreases below $\chi_c$, the aspect ratio drops sharply (the overshoot is due to inertial effects) and finally saturates at a value below unity.

\section*{Discussion:}
We introduced a model of an elastic shell as a two-dimensional network made of two orthogonal systems of elastic filaments arranged along the longitudinal and the latitudinal circles on a sphere, and used it to study shape transitions of active shells that are swollen by hydrostatic pressure. Activity was implemented via local principal curvature-dependent modification of the elastic constants of the nonlinear springs that lie along the longitudinal and latitudinal circles. 
%Throughout this paper we used the term ``curvature'' in the sense of polymer physics \cite{ref_4} but since our springs and expressed it in terms of the local bending angles of a freely-jointed extensible filament (bending rigidity is neglected in the current version of the model). 
We also allowed for active modification of the hydrostatic pressure inside the shell. The activation of local network stresses depends on the volume/area and on the shape of the elastic shell (through their effects on the curvatures of the springs) and reacts back on them, changing the area and shape that produced the activation in the first place. The exploration of the resulting feedback between morphology and elasticity was the main goal of our study.  

We examined several models of active deformation. We began by introducing hydrostatic pressure that swells and deforms the original sphere into a spheroid (a sphere flattened at the poles). This defined our initial state. In model $1$, activity was implemented by irreversibly reducing the spring constants of springs whose curvature $\chi$ fell below a critical value $\chi_c$. In model $2$ activation was assumed to be reversible, i.e. if the curvature of an activated spring increased above $\chi_c$, its spring constant was instantaneously restored to its equilibrium value.  Since the critical curvature was chosen to lie below the bottom of the curvature spectrum of latitudinal springs (see Figure \ref{fig_2}), we expected that only longitudinal springs will be activated and become stretched, and that an elongated ellipsoidal shape would result.  Surprisingly, we found that the above procedure has led to oblate spheroidal shapes. Upon some reflection we realized that since in models $1$ and $2$ activation of a spring was defined as reduction of its elastic constant, its primary effect was to increase the volume/area of the elastic shell. The resulting decrease of the local curvatures drove the curvature of some latitudinal (as well as longitudinal) springs below the critical value. As these latitudinal springs became activated, their elastic constants were reduced, giving rise to oblate spheroidal shapes. We found that in order to obtain elongated shapes (prolate ellipsoids), the area increase and the resulting activation of latitudinal springs, had to be controlled by reducing the hydrostatic pressure. 
%In living organisms this can be attained through active control of osmotic pressure differences between the inside and the outside of the shell \cite{ref_5}.
The exploration of the pressure dependence of the steady state  shape of the elastic shell in the presence of active excitations, showed that in both models $1$ and $2$ there was a critical pressure at which a discontinuous transition from an elongated ellipsoid (at $p<p_c$) to a spheroid (at $p>p_c$) took place. The existence of this discontinuous shape and area transition suggests that it can be used as a switch: a small change of the hydrostatic pressure (e.g., by active control of osmotic pressure) can induce dramatic change of the size and the shape of the elastic shell. Interestingly, a striking reversal of morphogenesis from elongated mature hydra to a spheroidal morphology upon application of electric fields has been recently observed \cite{braun2019electric}.

We also explored a model (model $3$) in which all springs were actively excited: those with curvature lower than critical were softened while those above critical were stiffened. As the network deformed, springs continuously adjusted their elastic constants according to whether their instantaneous curvature was larger or smaller than critical. In this model, complex reentrant behavior of the steady state shape with changing pressure was observed. In the low pressure limit, the shape was spheroidal (aspect ratio below unity). A nearly continuous transition to prolate ellipsoidal shape (aspect ratio above unity) with increasing pressure was observed, along with discontinuous jumps of area and aspect ratio at pressures $p_{c1}$ and $p_{c2}$ (note that since in all of our models the elastic constants of a group of springs change discontinuously each time their curvature crosses $\chi_c$, a step-like response of the area and shape of the network to pressure changes is expected). While surface area always increased with increasing pressure, non-monotonic dependence of the aspect ratio on pressure was observed in a limited pressure range. Similarly to models $1$ and $2$, further increase of the pressure resulted in a discontinuous transition from a prolate ellipsoid to a strongly inflated oblate spheroid at some critical value of  pressure, $p_{c3}$. 

The observation of discontinuous area and shape transitions between elongated ellipsoidal and spheroidal shapes upon change of pressure, is one of the important results of our work. Another important attribute of all the models we studied is that an absorbing state (we referred to it as ``steady state'') in which no further changes of shape and no further excitations of springs were possible, was always reached in our simulations.  While volume (area) transitions are ubiquitous in nature (e.g., the liquid-gas transition at constant pressure), shape transitions have been mostly observed in ``smart'' pre-designed materials such as swollen gels with inhomogeneous  network topology\cite{shahinpoor1998ionic,klein2007shaping,panyukov2015cross}.
In our model, the network topology is inhomogeneous as well: the structure at the poles is different from that in the bulk of the network. While this inhomogeneous structure is sufficient to generate the sphere-to-spheroid transition upon increase of the hydrostatic pressure even in the absence of active excitations, curvature-dependent active excitations of the springs have to be introduced in order to give rise to elongated ellipsoidal shapes. We would like to stress that while shape change is prescribed in most ``smart'' materials by programming a desired target metric, the curvature-dependent excitations in our network model lead to self-organized shape change, \emph{i.e. the target metric responds to the current deformation state}. Also, we note that while the discontinuous change in shape at a critical pressure resembles a first order phase transition, such phase transitions occur typically in thermodynamic (infinite) systems and appear continuous in simulations of finite systems. In fact, the origin of discontinuous shape transitions in our finite system can be traced back to the discrete spectrum of curvatures in our model. 

An important difference between our approach and that of previous studies  of active solids is that in our model activity is implemented via a change of the spring constants of the network and is therefore a scalar property; the directional character of the shape transitions can be traced back to the anisotropic topology of the elastic network that consists of cross-linked longitudinal and latitudinal springs. This  differs from models in which the coupling between elasticity and activity is achieved via active polar agents that act at the vertices of an elastic network \cite{baconnier2022selective}, or through the 
introduction of an anisotropic active stress coupled to local orientational order \cite{pismen2014spontaneous, maitra2019oriented}.  
%but, because it involves non-conservative forces, it may be related to ``odd elasticity'' recently introduced by Vittelli and coworkers \cite{ref_8}.
A corollary of the fact that activity is introduced by changing the parameters of the elastic energy of the network, is that the corresponding elastic forces are not conservative (i.e., can not be represented as gradients of potentials) \cite{osmanovic2021spatial}. Therefore, the connection between our model and the standard continuum theory of elasticity \cite{ref_7} which is often assumed in theories of active solids \cite{pismen2014spontaneous, maitra2019oriented, zakharov2021modeling}, is not well-understood at present.

We would like to comment on some limitations of the present work. We introduced a particular geometrical construction of the elastic network, a choice that has enabled us to generate shape transitions from spheroids to elongated ellipsoids. Other regular (see, e.g., \cite {ref_1}) and random network structures were not explored in this work. We considered several models of active excitation of the elastic shell but, clearly, the number of possible models is very large. Future explorations of the model and of the parameter space to design spring networks optimized for a target response, such as shown for an ``allostery''-inspired long-range but local network response \cite{rocks2017designing} may be possible. Furthermore, we considered only the overdamped regime and did not carry out an exhaustive study of the effects of inertia and friction. 
%Another limitation of the present work is the neglect of fiber bending rigidity which dominates the elasticity of biopolymer networks; such effects can be easily incorporated into our model and we plan to explore them in the future. 

Finally, we would like to comment on the relevance of our work to morphogenesis in reconstituted hydra which provoked our interest in the more general problem of shape transitions in active elastic shells. While we took into account, albeit heuristically, the organization of the supracellular actomyosin fibers that run through the body of the hydra and act as muscles that stretch and compress the hydra tissue, we did not account for the fluctuations of calcium ions that activate these muscles \cite{agam2023hydra}. Still, we were able to reproduce some of the qualitative features of hydra morphogenesis such as the spheroid to elongated ellipsoid transition that is accompanied by steep increase of the surface area (as in our model $3$). This success notwithstanding, our work should not be considered as a phenomenological model of any concrete biological system.  Rather, it is a study of a minimal physical model which provides important insights about the mechanisms that underlie shape transformations in active solids, whether of artificial (smart materials) or biological origin.

\section*{Acknowledgement}
YR would like to thank David Kessler for helpful discussions, suggestions and Eli Sloutskin for critical comments on the manuscript. AM would like to acknowledge the hospitality of the Physics department of Bar-Ilan University where part of the work was done. AM would like to thank Yuval Garini and Biophysics, Nano-dynamics lab (Technion-IIT) where the rest of the work has been performed.

%\bibliographystyle{unsrt}
%\bibliography{ref.bib}


\section*{Supplementary material}

# Supplementary Information

# Shape Transitions in Network Model of Active Elastic Shells

Ajoy Maji[a], Kinjal Dasbiswas [b], Yitzhak Rabin[c]

[a]Faculty of Biomedical Engineering, Technion-Israel Institute of Technology, 32000 Haifa, Israel
[b] Department of Physics, University of California, Merced, Merced, CA 95343, USA
[c] Department of Physics and Institute of Nanotechnology and Advanced Materials, Bar-Ilan University, Ramat-Gan 5290002, Israel

# Virtual triangulation of the network:

The 364 quadrilaterals are virtually triangulated by finding the center of mass of each of the quadrilateral and virtually connecting the center of mass to each of the vertices of that quadrilateral.

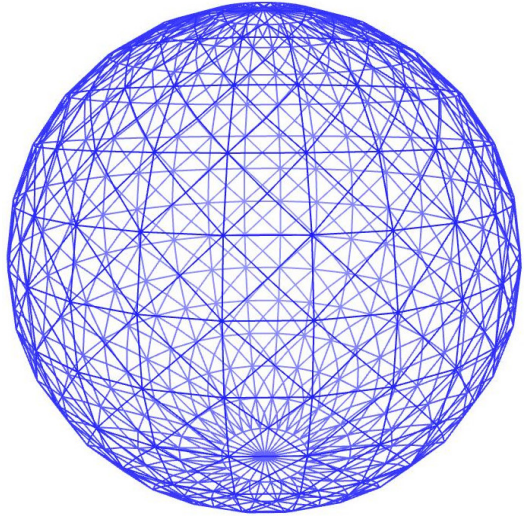

Fig. S 1: Snapshot of the virtually triangulated network

We introduce the hydrostatic force $\vec{F}_k^h$ on vertex k as the vector sum of the area of the triangles that meet at this vertex. $m = 28$ for the two poles, $m = 6$ for vertices that lie on the latitudinal circles next to the poles, and $m = 8$ for all other vertices.

$$\vec{F}_k^h = p \sum_{i=1}^{m} a_i \hat{n}_i \qquad (1)$$

Where $\vec{a}_i = a_i \hat{n}_i$ is the area vector of the $i^{th}$ triangle. The total area of the network is given by

$$\vec{A} = \sum_{i=1}^{1456+56} \vec{a}_i \qquad (2)$$

# Integration algorithm

$$\frac{d}{dt}(e^{\zeta t} \vec{v}_k) = e^{\zeta t} \vec{F}_k$$

Integrating both sides of this equation,

$$\int_{-\Delta/2}^{t} dt' \frac{d}{dt'}(e^{\zeta t'} \vec{v}_k) = \int_{-\Delta/2}^{t} dt' \, \vec{F}_k \, e^{\zeta t'} \qquad (3)$$

where $\Delta$ is the MD time step which we take to be much shorter than the viscous damping time (mass/(friction coefficient)).

Solving equation 3, we get the following expressions for updating the velocity and position of vertex $K$:

$$\vec{v}_k((n+\frac{1}{2})\Delta) = \vec{v}_k((n-\frac{1}{2})\Delta)e^{-\zeta\Delta} + e^{-\zeta\frac{\Delta}{2}} \Delta \, \vec{F}_k(\vec{r}_k(n\Delta)) \qquad (4)$$

$$\vec{v}_k(t) = \vec{v}_k(-\frac{\Delta}{2})e^{-\zeta(t+\Delta/2)} + e^{-\zeta t} \int_{-\Delta/2}^{t} dt' \vec{F}_k(t')e^{\zeta t'}$$

$$\implies \vec{v}_k(\tfrac{\Delta}{2}) = \vec{v}_k(-\tfrac{\Delta}{2})e^{-\zeta\Delta} + \Delta \vec{F}_k(\vec{r}_k(t=0))e^{-\zeta\frac{\Delta}{2}}$$

In general,

$$\vec{v}_k((n+\frac{1}{2})\Delta) = \vec{v}_k((n-\frac{1}{2})\Delta)e^{-\zeta\Delta} + e^{-\zeta\frac{\Delta}{2}} \Delta \, \vec{F}_k(\vec{r}_k(n\Delta)) \qquad (5)$$

where $n$ is the number of MD time steps.
The equation for the position can be written as,

$$\vec{r}_k(t) = \vec{r}_k(0) + \int_0^t \vec{v}_k(t')dt'$$

At $t = \Delta$,

$$\vec{r}_k(\Delta) = \vec{r}_k(0) + \Delta \, \vec{v}_k(\Delta/2)$$

In other words,

$$\vec{r}_k((n+1)\Delta) = \vec{r}_k(n\Delta) + \Delta \, \vec{v}_k \left((n+\frac{1}{2})\Delta\right) \qquad (6)$$

We are using equation 5 and 6 for updating velocities and positions of the vertices

# Representation of the angles between the tangent of the neighboring springs:

$\theta_i$ is the angle between the tangents of $i^{th}$ and $i-1^{th}$ spring. $\theta_{i+1}$ is the angle between the tangents of $i^{th}$ and $i+1^{th}$ spring.

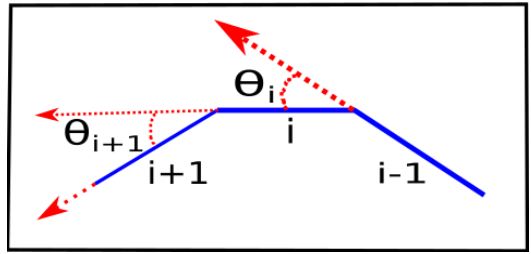

Fig. S 2: Representation of the angles between the tangents of the neighboring springs

# Steady state configuration of the network with $p_0 = 0.1$

The network takes an oblate spheroid shape in steady state with $p_0 = 0.1$, in the absence of active excitations. The distance between two poles (D) and the diameter of the equatorial plane (2R) are marked. The ratio of D/2R represents the aspect ratio of the shape.

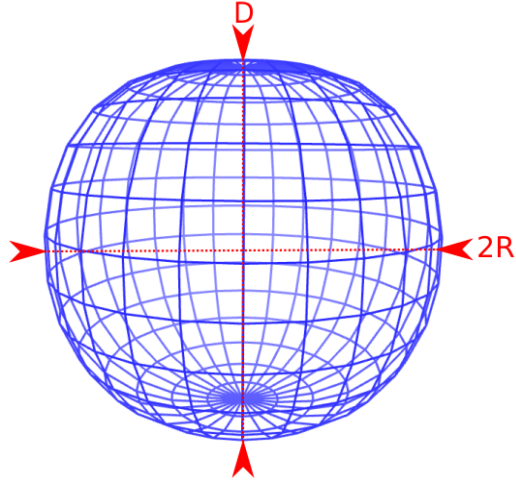

Fig. S 3: Steady state configuration of the network with $p_0 = 0.1$ before active excitations are switched on.

## Model 2. Curvature-dependent relaxation of activation: pressure-induced discontinuous transition from prolate to oblate ellipsoids

We begin again from the spheroidal configuration of the network at pressure $p_0 = 0.1$ (Figure S3). Just as in model 1 activity is switched on at $t = 500$ and the elastic constants of all springs whose instantaneous curvature is lower than $\chi_c = 0.05$ are reduced from 1 to 0.05. Simultaneously, the hydrostatic pressure is dropped from 0.1 to some lower value, $p$. If the curvature of a spring goes above $\chi_c$ in the course of dynamics, the spring constant of that spring returns to its equilibrium value $K_{eq} = 1$.

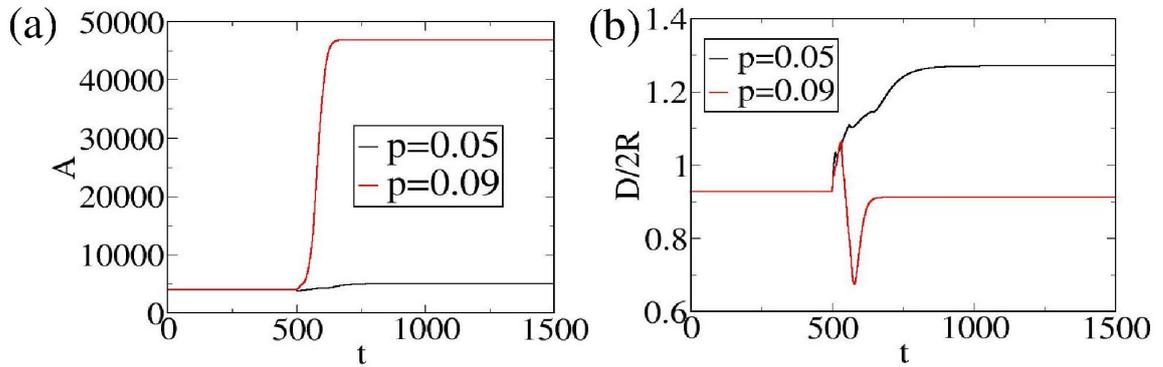

Fig. S 4: Model 2: (a) Area A and (b) aspect ratio $D/2R$ as a function of time for $p = 0.05$ (black) and $p = 0.09$ (red).

The area and the aspect ratio of model 2 are plotted as a function of time in Figures S4 (a) and (b) respectively, for $p = 0.05$ and $p = 0.09$. Slight differences between models 1 and 2 are observed at lower pressure, $p = 0.05$: in particular, the introduction of curvature-dependent relaxation in model 2 leads to lower area and aspect ratio compared to model 1 (Figure 3).

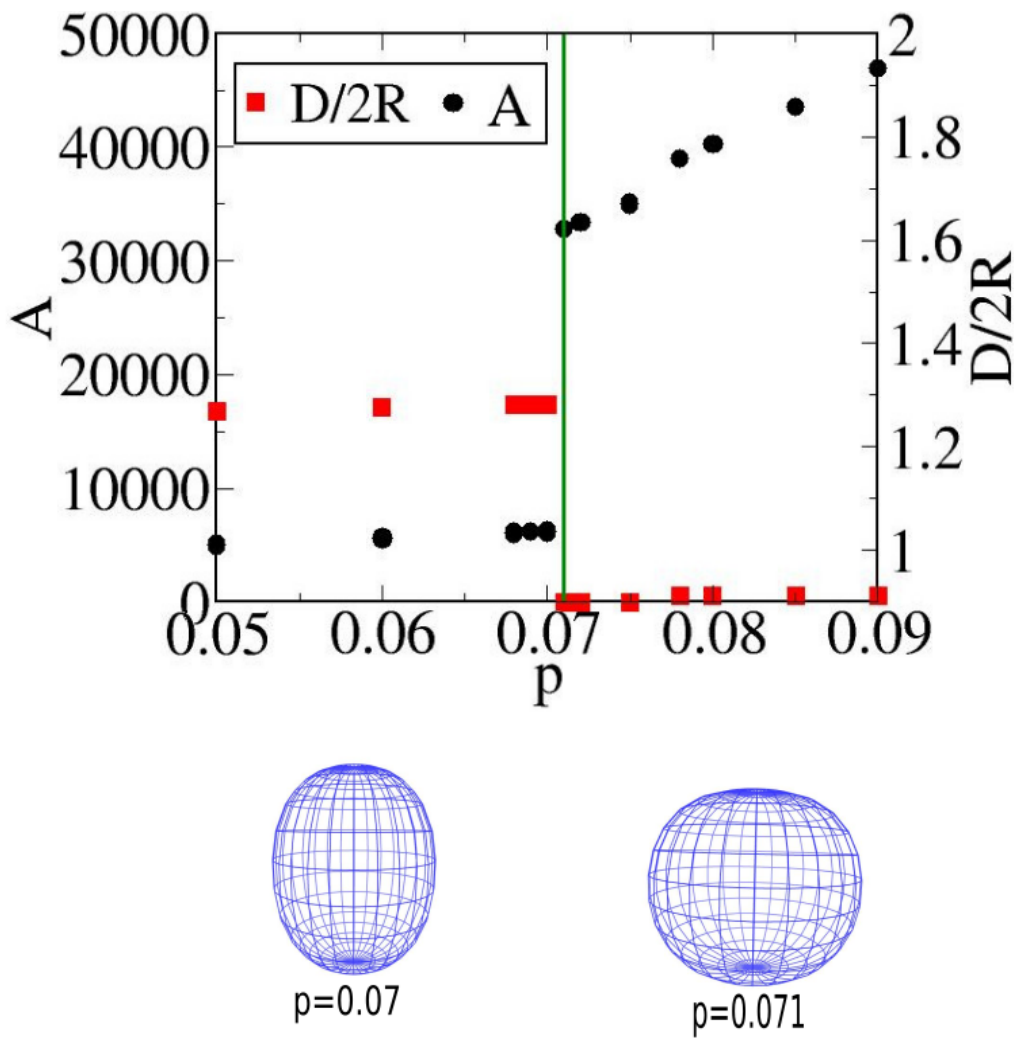

Fig. S 5: Model 2: area A and aspect ratio $D/2R$ are shown as a function of pressure. The location of the discontinuous transition is shown by the vertical broken green line. Snapshots of steady state configurations on both sides of the transition are presented in the lower panel (the snapshots are not to scale).

In Figure S5 we present the values of the area (black points) and the aspect ratio (red squares) in the final steady state of the actively excited network, for pressures in the range from 0.05 to 0.09, in the case of model 2. Similarly to model 1, a discontinuous transition from prolate to oblate ellipsoidal shape (accompanied by a jump of the total area of the network)

is observed between $p = 0.07$ and $p = 0.071$ (the dashed green vertical line in Figure S5). Snapshots of the steady state shapes of the system on both sides of the transition, at $p = 0.07$ and $p = 0.071$, are shown in the lower panel. Comparison of Figures 4 and S5 shows that the transition is shifted to slightly higher pressures and that the aspect ratios of the prolate ellipsoids for $p < p_c$ are lower in model 2 than in model 1.

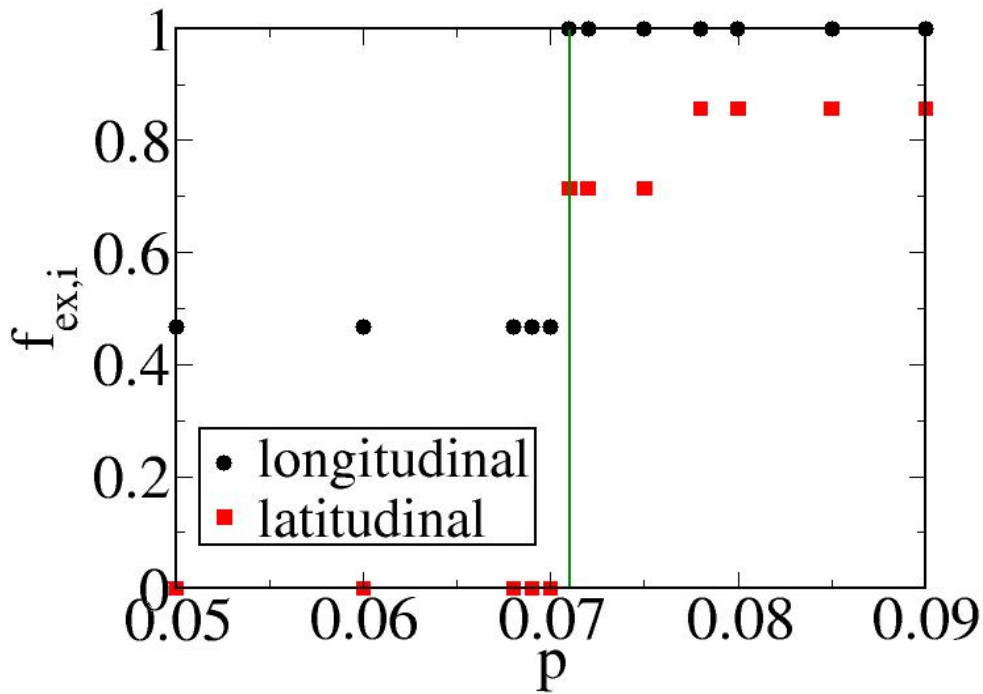

Fig. S 6: Model 2: Fractions of longitudinal and latitudinal activated springs in the excited steady state, as a function of pressure p.

In Figure S6 we present the fractions of activated longitudinal and latitudinal springs for different pressures in the range $0.05 - 0.09$. Comparison with Figure 6 shows that relaxation does not affect the size of the jump in the fraction of excited latitudinal springs but increases the discontinuity in the fraction of excited longitudinal springs across the transition.

# Model 3: Curvature Distributions

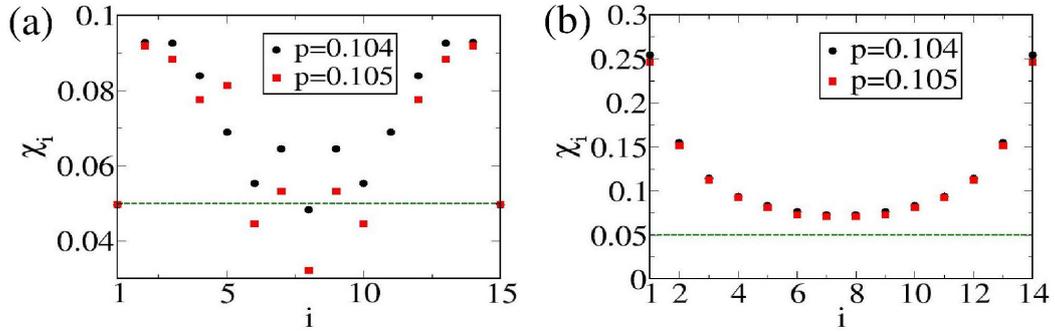

Fig. S 7: Model 3: Curvature distributions of (a) longitudinal and (b) latitudinal springs below and above $p_{c1}$, at $p = 0.104$ and $p = 0.105$, respectively.

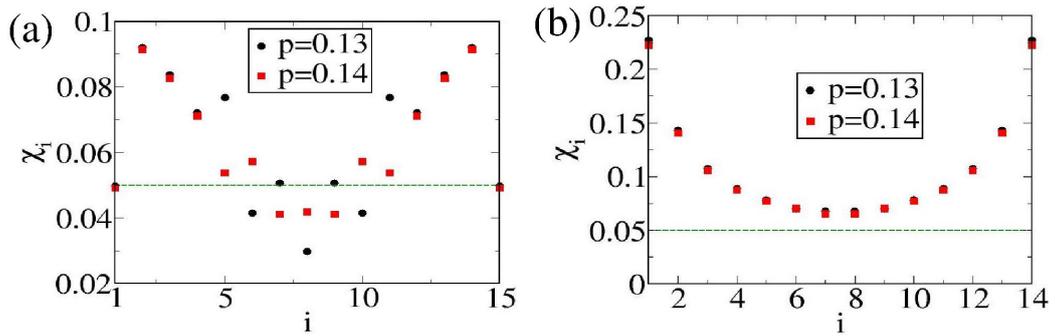

Fig. S 8: Model 3: Curvature distributions of (a) longitudinal and (b) latitudinal springs at $p = 0.13$ and $p = 0.14$.

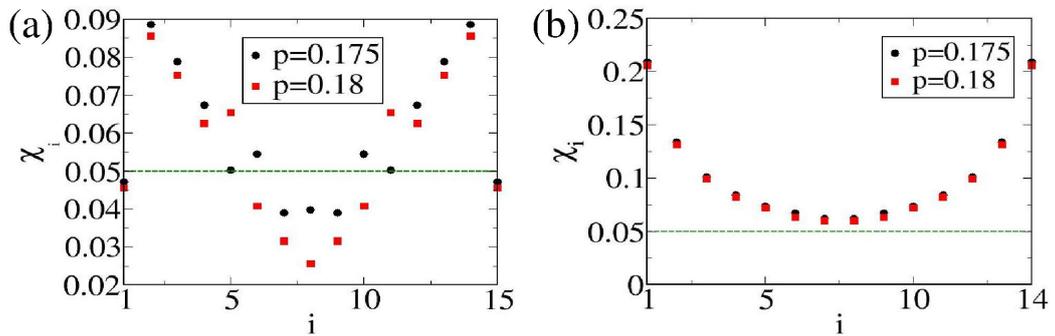

Fig. S 9: Model 3: Curvature distributions of (a) longitudinal and (b) latitudinal springs at $p = 0.175$ and $p = 0.18$.

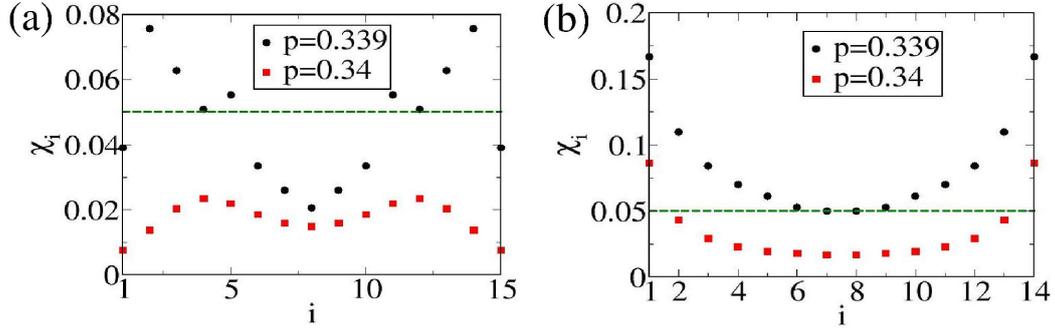

Fig. S 10: Model 3: Curvature distributions of (a) longitudinal and (b) latitudinal springs below and above $p_{c2}$, at $p = 0.339$ and $p = 0.34$, respectively.

# Movies:

**M1:** Model 1: Time evolution of the network to steady state with $p = 0.09$. Initially longitudinal springs are excited, but as swelling proceeds, latitudinal springs also get excited which leads to a runaway radial expansion of the network normal to the polar axis.

**M2:** Model 1: Time evolution of the network to steady state with $p = 0.05$. Swelling of the network under this comparatively lower $p$ value do not activate the latitudinal springs. The activation of the longitudinal springs leads to a prolate shape of the network.

**M3:** Model 1: The time evolution of the network to steady state for $p = 0.0653$, just before the transition to happen. Time evolution of the network at $p = 0.0653$ is very much similar to $p = 0.05$ (Movie M2) .

**M4:** Model 1: The time evolution of the network to steady state for $p = 0.0654$, just after the transition happens. Initially the activation of the longitudinal springs leads to a transient prolate shape of the network. But as the swelling progress, the latitudinal springs get activated and a dramatically swollen oblate spheroid results.

**M5:** Model 3: The time evolution of the network to steady state for $p = 0.02$. Since at t=0, the curvatures of most of the springs lie above $\chi_c$, the activation of the springs leads to contraction of the network, which fur-

10

ther increases the curvature of the springs. The shrinking of the area becomes a runaway process which ultimately relaxes to oblate spheroid.

**M6:** Model 3: The time evolution of the network to steady state for $p = 0.2$. Contraction of most of the activated springs is opposed by the increment of pressure. The elongation of the longitudinal springs leads to a prolate spheroid shape of the network.

**M7:** Model 3: The time evolution of the network to steady state for $p = 0.4$. At this high pressure value, the latitudinal springs also get activated along with longitudinal springs. This leads to an oblate spheroid shape of the network.


\end{document}